\documentclass[lettersize,journal]{IEEEtran}
\usepackage{amsmath,amsfonts}
\usepackage{algorithmic}
\usepackage{algorithm}
\usepackage{array}
\usepackage[caption=false,font=normalsize,labelfont=sf,textfont=sf]{subfig}
\usepackage{textcomp}
\usepackage{stfloats}
\usepackage{url}
\usepackage{verbatim}
\usepackage{graphicx}
\usepackage{cite}
\usepackage{amssymb}
\usepackage[colorlinks=true,citecolor=blue,linkcolor=blue,urlcolor=blue]{hyperref}

\newcommand{\bcite}[1]{{\color{blue}\cite{#1}}}

\newcommand{\beqref}[1]{{\color{blue}\eqref{#1}}}

\begin{document}

\title{Sum Rate Maximization for Movable Antenna Enabled Uplink NOMA}

\author{Nianzu Li,~Peiran Wu,~\IEEEmembership{Member,~IEEE},~Boyu Ning,~\IEEEmembership{Member,~IEEE} and Lipeng Zhu,~\IEEEmembership{Member,~IEEE}
\thanks{This work was supported in part by the National Natural Science Foundation of China under Grant 62171486 and Grant U2001213. The associate editor coordinating the review of this article and approving it for publication was Glauber Brante. \textit{(Corresponding author: Peiran Wu.)}}        

\thanks{Nianzu Li and Peiran Wu are with School of Electronics and Information Technology, Sun Yat-sen University, Guangzhou 510006, China (e-mails: linz5@mail2.sysu.edu.cn; wupr3@mail.sysu.edu.cn). 
	Boyu Ning is with National Key Laboratory
	of Science and Technology on Communications, University of Electronic
	Science and Technology of China, Chengdu 611731, China (e-mail: boydning@outlook.com). 
	Lipeng Zhu is with the Department of Electrical and Computer
	Engineering, National University of Singapore, Singapore 117583 (e-mail: zhulp@nus.edu.sg).}}        



\maketitle

\begin{abstract}
Movable antenna (MA) has been recently proposed as a promising candidate technology for the next generation wireless communication systems due to its significant capability of reconfiguring wireless channels via antenna movement. In this letter, we study an MA-enabled uplink non-orthogonal multiple access (NOMA) system, where each user is equipped with a single MA. Our objective is to maximize the users' sum rate by jointly optimizing the MAs' positions, the decoding order and the power control. To solve this non-convex problem, we equivalently transform it into two tractable subproblems. First, we use the successive convex approximation (SCA) to find a locally optimal solution for the antenna position optimization subproblem. Next, we derive the closed-form optimal solution of the decoding order and power control subproblem. Numerical results show that our proposed MA-enabled NOMA system can significantly enhance the sum rate compared to fixed-position antenna (FPA) systems and orthogonal multiple access (OMA) systems. 

\end{abstract}


\section{Introduction}
\IEEEPARstart{H}{igher} capacity has always been the trend in the evolution of next generation wireless communication systems due to the ever rising demand of high data rates and upsurge in growth of mobile devices. However, conventional fixed-position antennas (FPAs) have certain limitations on enhancing the channel capacity because they cannot fully exploit the spatial variation of
wireless channels between the transmit and receive antennas\bcite{ref1}. To overcome this problem, movable antenna (MA) technology has been proposed for exploring more degrees of freedom (DoFs) in the spatial domain\bcite{ref2},\bcite{ref16},\bcite{ref3}. Different from FPAs, MAs can be smartly moved within a confined region by connecting to the radio frequency (RF) chains via flexible cables. Thus, by adjusting MAs at positions with more favorable channel conditions, the wireless channels are reconfigured to achieve higher capacity.

Existing works have demonstrated the superiority of MAs in improving the system performance of wireless applications\bcite{ref3},\bcite{ref4},\bcite{ref5},\bcite{ref19},\bcite{ref6},\bcite{ref7},\bcite{ref20},\bcite{ref8},\bcite{ref9},\bcite{ref11},\bcite{ref17}. The hardware architecture, field-response based channel models, and performance analysis of MA systems were presented in\bcite{ref3}. The channel capacity of the MA-aided multiple-input multiple-output (MIMO) system was maximized in\bcite{ref4} by jointly optimizing the positions of MAs and the transmit covariance matrix. The designs of the MA array-enhanced secure wireless transmission systems were investigated in\bcite{ref5},\bcite{ref19}. A novel compressed sensing-based channel estimation method for MA-aided communication systems was proposed in\bcite{ref6}. The employment of MAs for wireless communications with coordinate multi-point (CoMP) reception was investigated in\bcite{ref7}. The utility of MA assistance for the multiple-input single-output (MISO) interference channel was investigated in\bcite{ref20}. Furthermore, exploiting MAs for multi-user communication scenarios was investigated in\bcite{ref8},\bcite{ref9},\bcite{ref11},\bcite{ref17}, where the data transmission is based on the space-division multiple access (SDMA). To the best of our knowledge, the study of MA-enabled non-orthogonal multiple access (NOMA) is still an open problem, which thus motivates our work.

In this letter, we study the application of MAs to enhance an uplink NOMA system, where each user is equipped with an MA to transmit signals to the base station (BS). Our objective is to maximize the users' sum rate, by jointly optimizing the positions of MAs, the decoding order and the power control. To handle this non-convex problem, we decouple it into two tractable subproblems. Specifically, we leverage the successive convex approximation (SCA) to solve the antenna position optimization subproblem, and then derive the closed-form optimal solution of the decoding order and power control subproblem. Finally, numerical results are presented, showing that the proposed MA-aided NOMA system greatly improves the sum rate compared to conventional FPA systems as well as orthogonal multiple access (OMA) systems.

\section{System model and problem formulation}
\subsection{System Model}
\begin{figure}[t]
	\centering
	\includegraphics[width=0.49\textwidth]{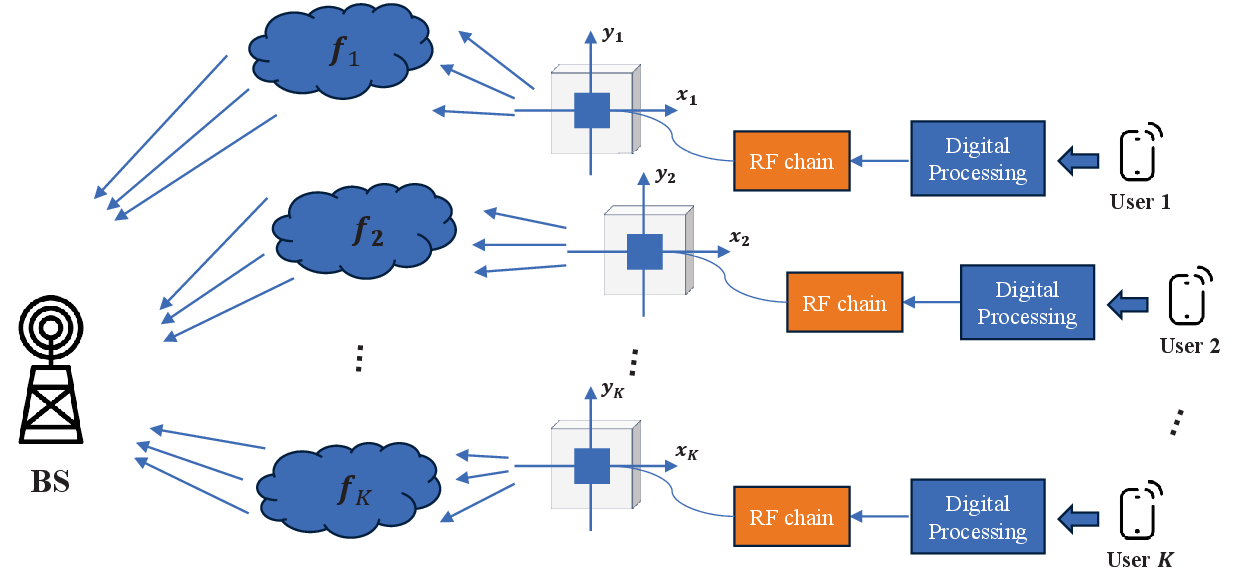}
	\caption{Illustration of the MA-enabled uplink NOMA system.}
	\label{system_model}
\end{figure}

As shown in Fig. \ref{system_model}, we consider an uplink NOMA system, where each user is equipped with an MA to transmit signals to the single-antenna BS\footnote{In practice, a multi-antenna BS usually synthesizes a beam to guarantee the coverage of a specific cluster, which can thus be equivalently regarded as a single-antenna BS.}. Each user's MA is connected to the RF chain via a flexible cable and its position can be flexibly moved in an $A\times A$ square area. Since the BS serves $K$ users simultaneously, the received signal is expressed as
\begin{equation}
	y=\sum_{k=1}^{K}h_k\sqrt{P_k}s_k+n,
\end{equation}
where $s_k$ denotes the transmitted signal from user $k$ with unit power, $P_k$ denotes the corresponding transmit power with the constraint of maximum value $P_{\mathrm{max}}$, $h_k$ denotes the channel response and $n$ denotes the additive white Gaussian noise with zero mean and variance $\sigma^2$. 

In this letter, we consider the field-response based channel model\bcite{ref3}, where the channel response is the superposition of complex coefficients of multiple channel paths. The positions of $K$ MAs are represented by the Cartesian coordinates, i.e., $\boldsymbol{z}_k=[x_k,y_k]^T\in\mathbb{R}^{2\times 1},k=1,\cdots,K$. The total number of transmit paths for the $k$th user-BS link is denoted as $L_k$, the elevation and azimuth angles of departure (AoDs) for the $p$th transmit path from user $k$ to the BS are denoted as $\theta_k^p\in[0,\pi]$ and $\phi_k^p\in[0,\pi]$, respectively. Therefore, the signal propagation distance difference of the $p$th transmit path  for user $k$ between position $\boldsymbol{z}_k$ and the original point $\boldsymbol{o}_k=[0,0]^T$ is 
\begin{equation}
	\rho_k^p(\boldsymbol{z}_k)=x_k\sin\theta_k^p\cos\phi_k^p+y_k\cos\theta_k^p.
\end{equation}
This indicates that the channel coefficient of the $p$th transmit path for user $k$ has a $2\pi\rho_k^p(\boldsymbol{z}_k)/\lambda$ phase difference over the original point $\boldsymbol{o}_k$, where $\lambda$ denotes the carrier wavelength. As a result, the transmit field-response vector (FRV) between user $k$ and the BS is given by\bcite{ref3}
\begin{equation}
	\boldsymbol{g}_k(\boldsymbol{z}_k)=\left[e^{j\frac{2\pi}{\lambda}\rho_k^1(\boldsymbol{z}_k)},e^{j\frac{2\pi}{\lambda}\rho_k^2(\boldsymbol{z}_k)},\cdots,e^{j\frac{2\pi}{\lambda}\rho_k^{L_k}(\boldsymbol{z}_k)}\right]^T.
\end{equation}
Finally, the channel response between user $k$ and the BS can be expressed as
\begin{equation}
	h_k=\boldsymbol{f}_k^H\boldsymbol{g}_k(\boldsymbol{z}_k),
\end{equation}
where $\boldsymbol{f}_k=[f_{k,1},f_{k,2},\cdots,f_{k,L_k}]^T$ denotes the $k$th user's path-response vector (PRV) at the BS. Therefore, by properly adjusting $\boldsymbol{z}_k$ based on the field-response information (FRI)\footnote{To implement antenna position optimization, the field-response information, including the AoDs and the coefficients of multiple channel paths, needs to be obtained. This can be realized by moving the MA to
a finite number of locations for channel measurement and using the compressed sensing-based sparse signal recovery algorithm for channel estimation\bcite{ref6}.}, a more favorable channel condition can be obtained to achieve a significant capacity gain.

According to the NOMA protocol, the BS employs successive interference cancelation (SIC) to recover the original signal. To this end, a decoding order should be assigned by the BS. Let $\pi_k$ denote the decoding order of user
$k$. For instance, if $\pi_k=n$, then the signal of user $k$ is the $n$th one to be decoded at the BS.
Thus, the signal-to-interference-plus-noise ratio (SINR) of user $k$ is given by
\begin{equation}
	\gamma_k=\frac{\left|\boldsymbol{f}_k^H\boldsymbol{g}_k(\boldsymbol{z}_k)\right|^2P_k}{\sum_{\pi_i>\pi_k}\left|\boldsymbol{f}_i^H\boldsymbol{g}_i(\boldsymbol{z}_i)\right|^2P_i+\sigma^2},
\end{equation}
and the achievable rate of user $k$ is given by
\begin{equation}
	R_k=\log_2\left(1+\gamma_k\right).
\end{equation}

\subsection{Problem Formulation}
In this letter, we aim to maximize the sum rate of all users  by jointly optimizing the positions of MAs, the decoding order of $K$ users and the transmit power, subject to a minimum rate requirement for each user. Let $\tilde{\boldsymbol{z}}=\left[\boldsymbol{z}_1^T,\boldsymbol{z}_2^T,\cdots,\boldsymbol{z}_K^T\right]^T$ denote the MA positioning vector, $\boldsymbol{\pi}=[\pi_1,\pi_2,\cdots,\pi_K]^T$ denote the decoding order vector, and $\boldsymbol{P}=\left[P_1,P_2,\cdots,P_K\right]^T$ denote the transmit power vector. Accordingly, the optimization problem can be formulated as
\begin{subequations}
	\label{eq7}
	\begin{align}
		\max_{\tilde{\boldsymbol{z}},\boldsymbol{\pi},\boldsymbol{P}}& \quad \sum_{k=1}^{K}\log_2\left(1+\gamma_k\right)\label{eq7a}\\
		\mathrm{s.t.}& \quad 
		\log_2\left(1+\gamma_k\right)\geq R_k^{\mathrm{min}},~k=1,\cdots,K,\label{eq7b}\\
		& \quad \boldsymbol{z}_k\in\mathcal{C}_k,~k=1,\cdots,K,\label{eq7c}\\
		& \quad 0\leq P_k\leq P_\mathrm{max},~k=1,\cdots,K,\label{eq7d}\\
		& \quad \boldsymbol{\pi} \in \boldsymbol{\Pi},\label{eq7e}
	\end{align}
\end{subequations}
where $R_k^\mathrm{min}$ is the minimum rate requirement for user $k$, $\mathcal{C}_k$ is the square moving region of the $k$th MA with size $[-A/2,A/2]\times [-A/2,A/2]$, and $\boldsymbol{\Pi}$ denotes the set of all possible decoding orders. Constraint \beqref{eq7b} indicates that the achievable rate of each user should not be smaller than its minimum requirement. Constraint \beqref{eq7c} confines the moving area of the MA for each user and constraint \beqref{eq7d} confines the transmit power of each user.

\section{Proposed solution}
Problem \beqref{eq7} is non-trivial because the sum rate is highly non-concave with respect to the positions of MAs, and constraint \beqref{eq7b} is also non-convex. Besides, the coupling among the optimization variables makes it more intractable. To solve this problem, we first re-express the objective function in a more concise form as
\begin{align}
	\label{eq8}
	R_\mathrm{sum}&=\sum_{k=1}^{K}\log_2\left(1+\tfrac{\left|\boldsymbol{f}_k^H\boldsymbol{g}_k(\boldsymbol{z}_k)\right|^2P_k}{\sum_{\pi_i>\pi_k}\left|\boldsymbol{f}_i^H\boldsymbol{g}_i(\boldsymbol{z}_i)\right|^2P_i+\sigma^2}\right)\notag\\
	&=\log_2\left(1+\tfrac{\sum_{k=1}^{K}\left|\boldsymbol{f}_k^H\boldsymbol{g}_k(\boldsymbol{z}_k)\right|^2P_k}{\sigma^2}\right).
\end{align} 
Then, we present the following proposition to transform problem \beqref{eq7} into two tractable subproblems.

\textit{Proposition 1}: The optimal solution of problem \beqref{eq7} can be obtained by solving two subproblems in sequence, as described in the following two steps.

Step 1: Obtain the optimal $\tilde{\boldsymbol{z}}$ by maximizing the channel gain of each user independently, i.e.,
\begin{align}
	\label{eq9}
	\max_{\boldsymbol{z}_k}&\quad|h_k|^2\\
	\mathrm{s.t.}& \quad \text{\beqref{eq7c}}.\notag
\end{align}

Step 2: Obtain the optimal $\boldsymbol{\pi}$ and $\boldsymbol{P}$ by solving the primal problem  with given $\tilde{\boldsymbol{z}}$, i.e.,
\begin{align}
	\max_{\boldsymbol{\pi},\boldsymbol{P}}& \quad R_\mathrm{sum}\label{eq10}\\
	\mathrm{s.t.}& \quad \text{\beqref{eq7b},~\beqref{eq7d},~\beqref{eq7e}}.\notag
\end{align}

\textit{Proof}: See Appendix A.$\hfill\blacksquare$

As we can see, problem \beqref{eq9} is still a non-convex optimization problem. To handle this issue, we use the SCA technique to relax it and present the locally optimal solution in Section III-A. Besides, given the obtained MA positioning vector $\tilde{\boldsymbol{z}}$, problem \beqref{eq10} is tractable and the optimal solution is derived in closed forms, which is presented in Section III-B.

\subsection{Antenna Position Optimization}
First, the objective function in problem \beqref{eq9} can be re-expressed as
\begin{align}
	\label{eq11}
	F(\boldsymbol{z}_k)&\triangleq\left|\boldsymbol{f}_k^H\boldsymbol{g}_k(\boldsymbol{z}_k)\right|^2\notag\\
	&=\boldsymbol{g}_k(\boldsymbol{z}_k)^H\boldsymbol{f}_k\boldsymbol{f}_k^H\boldsymbol{g}_k(\boldsymbol{z}_k)\notag\\
	&\triangleq\boldsymbol{g}_k(\boldsymbol{z}_k)^H\boldsymbol{B}_k\boldsymbol{g}_k(\boldsymbol{z}_k),
\end{align}
where $\boldsymbol{B}_k$ is a positive definite matrix, defined as $\boldsymbol{B}_k\triangleq\boldsymbol{f}_k\boldsymbol{f}_k^H\in\mathbb{C}^{L_k\times L_k}$. However, the above expression is still non-concave with respect to $\boldsymbol{z}_k$. To tackle this problem, we use the SCA technique to relax it\bcite{ref13},\bcite{ref18}. Since \beqref{eq11} is a convex function with respect to $\boldsymbol{g}_k(\boldsymbol{z}_k)$, by applying the first-order Taylor expansion, the lower bound on $F(\boldsymbol{z}_k)$ at given local point $\boldsymbol{z}_k^i$ can be given by
\begin{align}
	\label{eq12}
	F(\boldsymbol{z}_k)\geq&2\mathrm{Re}\left\{\boldsymbol{g}_k(\boldsymbol{z}_k^i)^H\boldsymbol{B}_k\left(\boldsymbol{g}_k(\boldsymbol{z}_k)-\boldsymbol{g}_k(\boldsymbol{z}_k^i)\right)\right\}\notag\\
	&+\boldsymbol{g}_k(\boldsymbol{z}_k^i)^H\boldsymbol{B}_k\boldsymbol{g}_k(\boldsymbol{z}_k^i)\notag\\
	=&2\underbrace{\mathrm{Re}\left\{\boldsymbol{g}_k(\boldsymbol{z}_k^i)^H\boldsymbol{B}_k\boldsymbol{g}_k(\boldsymbol{z}_k)\right\}}_{\bar{F}(\boldsymbol{z}_k)}\notag\\
	&-\underbrace{\boldsymbol{g}_k(\boldsymbol{z}_k^i)^H\boldsymbol{B}_k\boldsymbol{g}_k(\boldsymbol{z}_k^i)}_{\text{constant1}}.
\end{align}
Note that $\bar{F}(\boldsymbol{z}_k)$ is still non-concave over $\boldsymbol{z}_k$. Thus, we further adopt the second-order Taylor expansion to obtain a quadratic surrogate lower bound, given by
\begin{align}
	\label{eq13}
	\bar{F}(\boldsymbol{z}_k)\geq&\bar{F}(\boldsymbol{z}_k^i)+\nabla\bar{F}(\boldsymbol{z}_k^i)^T(\boldsymbol{z}_k-\boldsymbol{z}_k^i)\notag\\
	&-\frac{\delta_k}{2}(\boldsymbol{z}_k-\boldsymbol{z}_k^i)^T(\boldsymbol{z}_k-\boldsymbol{z}_k^i)\notag\\
	=&\underbrace{-\frac{\delta_k}{2}\boldsymbol{z}_k^T\boldsymbol{z}_k+\left(\nabla\bar{F}(\boldsymbol{z}_k^i)+\delta_k\boldsymbol{z}_k^i\right)^T\boldsymbol{z}_k}_{\tilde{F}(\boldsymbol{z}_k)}\notag\\
	&+\underbrace{\bar{F}(\boldsymbol{z}_k^i)-\left(\nabla\bar{F}(\boldsymbol{z}_k^i)+\frac{\delta_k}{2}\boldsymbol{z}_k^i\right)^T\boldsymbol{z}_k^i}_{\text{constant2}},
\end{align}
where $\nabla\bar{F}(\boldsymbol{z}_k)\in\mathbb{R}^{2}$ and $\nabla^2\bar{F}(\boldsymbol{z}_k)\in\mathbb{R}^{2\times2}$ denote the gradient vector and Hessian matrix of $\bar{F}(\boldsymbol{z}_k)$ over $\boldsymbol{z}_k$, respectively, $\delta_k$ is a positive number satisfying $\delta_k\boldsymbol{I}_2\succeq\nabla^2\bar{F}(\boldsymbol{z}_k)$, and the expressions of $\nabla\bar{F}(\boldsymbol{z}_k)$ and $\delta_k$ are provided in Appendix B. Therefore, in the $i$th iteration of SCA, the optimization problem is expressed as
\begin{subequations}
	\label{eq14}
	\begin{align}
		\max_{\boldsymbol{z}_k}& \quad \tilde{F}(\boldsymbol{z}_k)\\
		\mathrm{s.t.}& \quad \boldsymbol{z}_k\in\mathcal{C}_k,~\forall k\label{eq14b}.
	\end{align}
\end{subequations}
It is clear that problem \beqref{eq14} is a convex problem and the optimal solution can be obtained by using the CVX toolbox. Thus, in each iteration of SCA, we can obtain the optimal solution $\boldsymbol{z}_{k,i+1}^\star$ of problem \beqref{eq14} and set it as $\boldsymbol{z}_k^{i+1}$ for the next iteration. Finally, the iterative process of SCA terminates until the objective of problem \beqref{eq9} converges to a prescribed accuracy.

\subsection{Decoding Order and Power Control}
Since the objective of problem \beqref{eq10} is a logarithmic function, which is monotonically increasing with respect to $P_k$, problem \beqref{eq10} is equivalent to
\begin{subequations}
	\label{eq15}
	\begin{align}
		\quad \max_{\boldsymbol{\pi},\boldsymbol{P}}& \quad \sum_{k=1}^{K}\left|h_k\right|^2P_k\\
		\mathrm{s.t.}& \quad 
		\left|h_k\right|^2P_k\geq\alpha_k\left(\sum_{\pi_i>\pi_k}\left|h_i\right|^2P_i+\sigma^2\right),~\forall k,\\
		& \quad 0\leq P_k\leq P_\mathrm{max},~\forall k, \\
		& \quad \boldsymbol{\pi} \in \boldsymbol{\Pi},
	\end{align}
\end{subequations}
where $\alpha_k$ is a positive number defined as $\alpha_k\triangleq2^{R_k^{\min}}-1$. Note that problem \beqref{eq15} is a traditional joint optimization problem in an uplink NOMA system and the optimal solution is provided in the following proposition.

\textit{Proposition 2}: The optimal decoding order $\boldsymbol{\pi}^{\star}$ is the decreasing order of $|h_k|^2(1+\frac{1}{\alpha_k})$. For user 1, the optimal power control variable is $P_1^{\star}=P_{\max}$, while for user $k$ ($k\geq2$), the optimal power control variable is
\begin{equation}
	P_k^{\star}=
	\begin{cases}
		\min\{P_{\max},b_k\};&{\text{if}}\ P_i^{\star}=P_{\max},\forall i< k, \\
		c_k;&{\text{otherwise}},
	\end{cases}
\end{equation}
where
\begin{align}
	b_k=\min\Big\{\frac{|h_i|^2P_{\max}}{\alpha_i|h_k|^2}-\sum_{q=i+1}^{k-1}\frac{|h_q|^2P_{\max}}{|h_k|^2}-\notag\\
	\sum_{j=k+1}^{K}\frac{|h_j|^2c_j}{|h_k|^2}-\frac{\sigma^2}{|h_k|^2},1\leq i \leq k-1 \Big\}
\end{align}
and
\begin{equation}
	c_k=\frac{\sigma^2\alpha_k}{|h_k|^2}\prod_{i=k+1}^{K}(\alpha_i+1).
\end{equation}

\textit{Proof}: The proof can be found in\bcite{ref15}.$\hfill\blacksquare$

Therefore, we can obtain the closed-form optimal decoding order vector and transmit power vector of problem \beqref{eq10}.

\subsection{Overall Algorithm}
Based on the above two subproblems, the overall algorithm for solving problem \beqref{eq7} is summarized in Algorithm \ref{alg1}. 

\textit{Complexity Analysis}: In the optimization of $\boldsymbol{z}_k$, the
computational complexity is $\mathcal{O}(NL_k^2+N\ln(1/\beta))$, where $N$ denotes the iteration number of SCA and $\beta$ denotes the accuracy of the interior-point method\bcite{ref4}. The complexity of calculating $\boldsymbol{\pi}$ and $\boldsymbol{P}$ is $\mathcal{O}(K^2)$. Denoting $L$ as the maximum value of $\{L_{k}\}_{k=1}^K$, the overall complexity of Algorithm \ref{alg1} is
\begin{equation}
	\mathcal{O}\left(KNL^2+KN\ln(1/\beta)+K^2\right).
\end{equation}
\begin{algorithm}[H]
	\caption{Proposed Algorithm for solving Problem \beqref{eq7}.}
	\begin{algorithmic}[1]
		\STATE Initialize: $\{\boldsymbol{z}_k^0\}_{k=1}^K$
		\STATE \textbf{for} $k=1\rightarrow K$ \textbf{do}
		\STATE \hspace{0.25cm} $i = 0$.
		\STATE \hspace{0.25cm} \textbf{repeat}
		\STATE \hspace{0.25cm} Calculate $\nabla\bar{F}(\boldsymbol{z}_k^i)$ and $\delta_k$ via \beqref{eq21} and \beqref{eq22}, \\
		\hspace{0.25cm} respectively.
		\STATE \hspace{0.25cm} Obtain $\boldsymbol{z}_k^{i+1}$ by solving problem \beqref{eq14}.
		\STATE \hspace{0.25cm} $i = i+1$.
		 \STATE \hspace{0.25cm} \textbf{until} Increase of $|h_k|^2$ is below a threshold $\xi>0$.
		\STATE \textbf{end for}
		\STATE Obtain $\boldsymbol{\pi}$ and $\boldsymbol{P}$ via Proposition 2.
		\STATE \textbf{return} $\{\boldsymbol{z}_k\}_{k=1}^K$, $\boldsymbol{\pi}$ and $\boldsymbol{P}$.
	\end{algorithmic}
	\label{alg1} 
\end{algorithm}

\section{Numerical results}
\begin{figure}[t]
	\centering
	\includegraphics[width=0.46\textwidth]{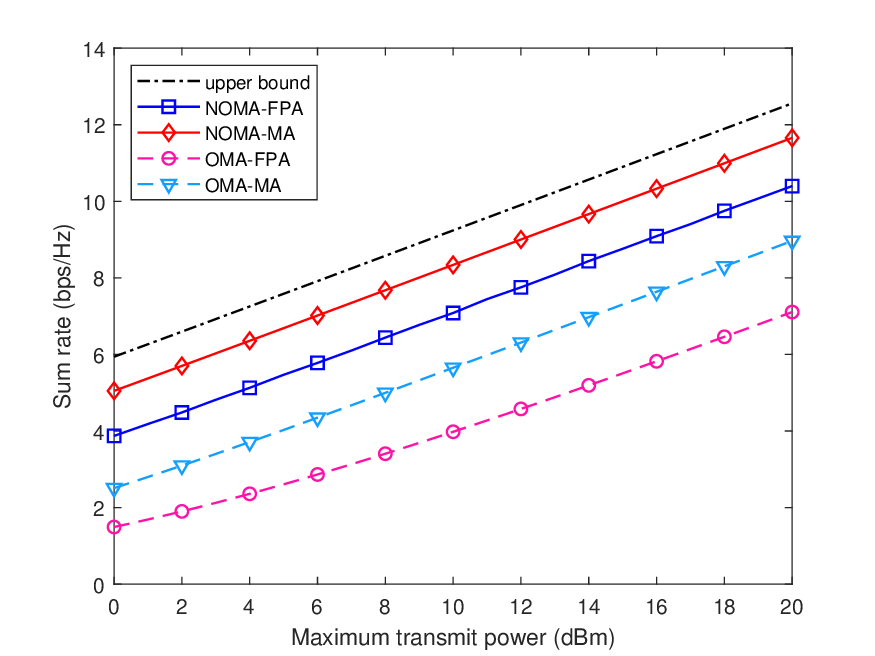}
	\caption{Sum rates versus the maximum transmit power $P_{\max}$ under different schemes, where $L=5,K=6$ and $R_k^{\min}=0.25$ bps/Hz.}
	\label{sum_rate_vs_Pmax}
\end{figure}
In this section, numerical results are provided to evaluate the performance of our proposed MA-enabled uplink NOMA system. Specifically, we consider the geometry channel model\bcite{ref3},\bcite{ref11}, where the numbers of transmit paths for all users are the same, i.e., $L_k=L,1\leq k\leq K$. The PRV for each user is modeled as a circularly symmetric complex Gaussian random vector with independent and identically distributed (i.i.d.) elements, i.e., $f_{k,n}\sim\mathcal{CN}(0,d_k^{-\alpha}/L),1\leq n\leq L$, where $\alpha$ denotes the path loss exponent, $d_k$ denotes the distance from user $k$ to the BS. The elevation and azimuth AoDs of each user are assumed to be i.i.d. variables with the uniform distribution over $[0,\pi]$. The distances from each user to the BS are randomly generated from $80$ to $100$ m; the path loss exponent is set to $\alpha=3.9$; the average noise power is set to $\sigma^2=-80$ dBm; the side length of the moving regions for each MA is set to $A=2\lambda$. In our simulations, all the results are averaged over $10^3$ independent channel realizations.

In Fig. \ref{sum_rate_vs_Pmax}, we present the sum rate versus the maximum transmit power under different schemes where the number of transmit paths for each user is $L=5$, the number of users is $K=6$, and the minimum rate requirement for each user is $R_k^\mathrm{min}=0.25$ bps/Hz. For both ``OMA-FPA" and ``OMA-MA" schemes, it is assumed that the time/frequency resources are equally allocated to $K$ users. For fair comparison, the side length of antenna moving regions for the ``OMA-MA" scheme is set identical to that of the ``NOMA-MA" scheme and its optimization procedure follows steps 2--9 in Algorithm \ref{alg1}, where each user's channel gain is maximized independently. Besides, we also plot the upper bound of sum rates, based on
\begin{align}
	~~~~~R_\mathrm{sum}&=\log_2\left(1+\tfrac{\sum_{k=1}^{K}\left|\boldsymbol{f}_k^H\boldsymbol{g}_k(\boldsymbol{z}_k)\right|^2P_k}{\sigma^2}\right)\notag\\
	&\leq\log_2\left(1+\tfrac{\sum_{k=1}^{K}\left(\sum_{n=1}^{L}|f_{k,n}|\right)^2P_{\max}}{\sigma^2}\right).
\end{align}
From Fig. \ref{sum_rate_vs_Pmax}, it is shown that the sum rate achieved by our proposed MA-aided NOMA system is much larger than those achieved by the conventional FPA systems. This is because compared to FPAs, MAs can exploit the spatial diversity of the users' channels more efficiently via antenna movement. It can also be observed that there is a significant performance gap between NOMA and OMA systems. The reason is that OMA only shares orthogonal communication resources among users, while NOMA distinguishes users with different channel conditions in power domain and utilizes interference cancellation of SIC to enable better spectral efficiency. Furthermore, the sum rate of the proposed “NOMA-MA” scheme approaches the upper
bound the most closely, verifying the effectiveness of employing MAs in NOMA systems.

\begin{figure}[t]
	\centering
	\includegraphics[width=0.46\textwidth]{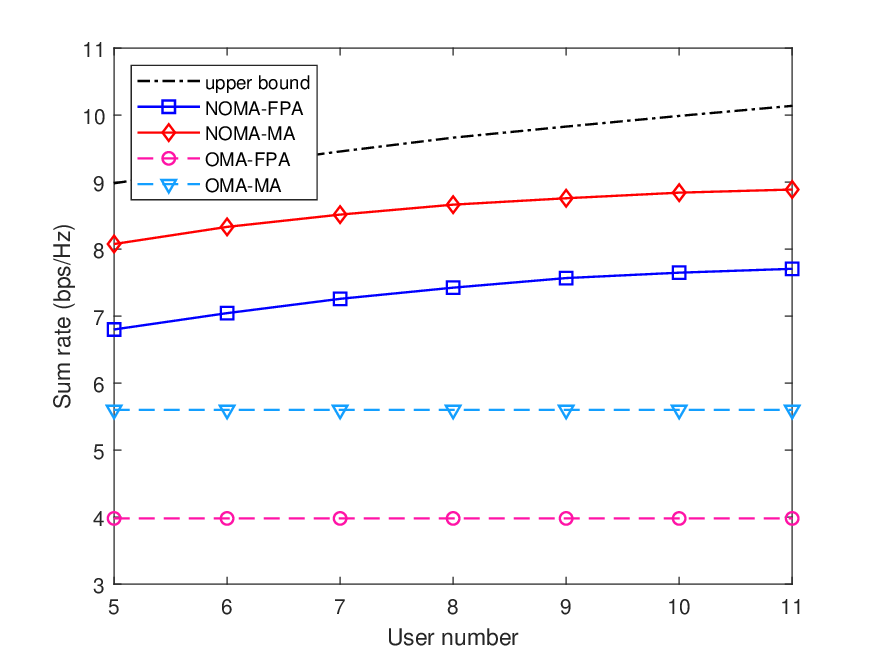}
	\caption{Sum rates versus the number of users $K$ under different schemes, where $L=5,P_{\max}=10$ dBm and $R_k^{\min}=0.25$ bps/Hz.}
	\label{sum_rate_vs_K}
\end{figure}
In Fig. \ref{sum_rate_vs_K}, we present the sum rate versus the number of users, where $L=5,P_{\max}=10$ dBm and $R_k^{\min}=0.25$ bps/Hz. For both NOMA or OMA systems, it is shown that the scheme with MAs achieves a significant gain over its counterpart with FPAs, which verifies the superiority of MAs. Besides, it is also found that the sum rate achieved by OMA systems remains unchanged for different user numbers since all users are served using orthogonal time/frequency resources. In contrast, the sum rate achieved by NOMA systems increases with the number of users due to the multiplexing gain in power domain. However, the increasing rate slows down when the number of users becomes large. This is because as the number of users grows, the interference among users increases such that the minimum rate requirements are difficult to be satisfied.
 


\section{Conclusion}
In this letter, we studied the sum rate maximization problem for a novel MA-enabled uplink NOMA system. The MA positioning vector, decoding order vector and transmit power vector of multiple users were jointly optimized to maximize the system sum rate, subject to the constraints of minimum rate requirement, maximum transmit power, and antenna moving region for each user. To solve this non-convex problem, we transformed it into two tractable subproblems and developed a low-complexity algorithm, in which the SCA technique was used to solve the antenna position optimization subproblem and the closed-form optimal solution was derived for the decoding order and power control subproblem. It was shown through numerical results that the proposed MA-aided NOMA system can outperform both FPA systems and OMA systems in terms of the achievable sum rate.



{\appendices
\section{Proof of Proposition 1}
From \beqref{eq8}, we know that $R_{\mathrm{sum}}$ is a monotonically increasing function with respect to $\sum_{k=1}^{K}|h_k|^2P_k$. Thus, maximizing $R_{\mathrm{sum}}$ is equivalent to maximizing $\sum_{k=1}^{K}|h_k|^2P_k$. Besides, in the objective function of problem \beqref{eq7} and constraint \beqref{eq7b}, only $|h_k|^2P_k$ is related with a specific $\{\boldsymbol{z}_k,P_k\}$. For a given feasible point $\{\boldsymbol{z}_{k,0},P_{k,0}\}$, we can always find another feasible point by solving problem \beqref{eq9} to maximize $|h_k|^2$ and then reduce the value of $P_{k,0}$ to keep the value of $\sum_{k=1}^{K}|h_k|^2P_k$ unchanged. Therefore, for any feasible solution of problem \beqref{eq7}, its objective value is no larger than that derived by solving problem \beqref{eq9} and \beqref{eq10}. This thus proves Proposition 1. 

\section{Expressions of $\nabla\bar{F}(\boldsymbol{z}_k)$ and $\delta_k$}
Define $\boldsymbol{b}\triangleq\boldsymbol{B}_k\boldsymbol{g}_k(\boldsymbol{z}_k^i)\in\mathbb{C}^{L_k\times1}$ and let $b_q=|b_q|e^{j\angle b_q}$ denote the $q$th entry of $\boldsymbol{b}$, with amplitude $|b_q|$ and phase $\angle b_q$. According to\bcite{ref4}, the gradient vector of $\bar{F}(\boldsymbol{z}_k)$ is given by
\begin{align}
	\label{eq21}
	\nabla\bar{F}(\boldsymbol{z}_k)&=\begin{bmatrix}
		\begin{aligned}
			&-\frac{2\pi}{\lambda}\sum_{q=1}^{L_k}|b_q|\sin\theta_k^q\cos\phi_k^q\sin(\kappa^q(\boldsymbol{z}_k))\\
			&-\frac{2\pi}{\lambda}\sum_{q=1}^{L_k}|b_q|\cos\theta_k^q\sin(\kappa^q(\boldsymbol{z}_k))
		\end{aligned}
	\end{bmatrix},
\end{align}
where $\kappa^q(\boldsymbol{z}_k)\triangleq2\pi\rho_k^q(\boldsymbol{z}_k)/\lambda-\angle b_q$, and the positive number $\delta_k$ is selected as

\begin{equation}
	\label{eq22}
	\delta_k=\frac{8\pi^2}{\lambda^2}\sum_{q=1}^{L_k}|b_q|,
\end{equation}
which satisfies $\delta_k\boldsymbol{I}_2\succeq\nabla^2\bar{F}(\boldsymbol{z}_k)$.

}

\bibliography{reference}
\bibliographystyle{IEEEtran}

\vfill
\end{document}